# Biexciton Initialization by Two-Photon Excitation in Site-Controlled Quantum Dots: the Complexity of the Antibinding State Case

Gediminas Juska,[1, a)] Iman Ranjbar Jahromi,[1] Francesco Mattana,[1] Simone Varo,[1] Valeria Dimastrodonato,[1] and Emanuele Pelucchi[1]

*Tyndall National Institute, University College Cork, Lee Maltings, Dyke Parade, T12R5CP, Cork, Ireland*

(Dated: 4 October 2020)

In this work, we present a biexciton state population in (111)B oriented site-controlled InGaAs quantum dots (QDs) by resonant two photon excitation. We show that the excited state recombines emitting highly pure single photon pairs entangled in polarization. The discussed cases herein are compelling due to the specific energetic structure of Pyramidal InGaAs QDs – an antibinding biexciton – a state with a positive binding energy. We demonstrate that resonant two-photon excitation of QDs with antibinding biexcitons can lead to a complex excitation-recombination scenario. We systematically observed that the resonant biexciton state population is competing with an acoustic-phonon assisted population of an exciton state. These findings show that under typical two-photon resonant excitation conditions deterministic biexciton state initialization can be compromised. This complication should be taken into account by the community members aiming to utilise similar epitaxial QDs with an antibinding biexciton.

Quantum dots (QDs) are sources of non-classical light – single and entangled photons – a resource which can be used to implement quantum information processing. Practically a QD has to be excited either optically or electrically to create excitonic complexes occupying the atomic-like energetic levels, which then spontaneously recombine emitting single photons. A specific excitonic complex composed of two pairs of electrons and holes with a net zero spin – a biexciton – recombines in cascaded events[1] emitting a pair of polarization-entangled photons[2]. It is a very attractive resource for quantum information processing, for example, it potentially allows for a simplification of quantum circuit designs due to a reduced number of entangling gates, or for innovative heralded schemes[3]. Historically, the typical approach to excite a QD to the biexciton state has been of non-resonant nature (optical or electrical excitation of the QD barriers) – a process which is incoherent, noisy and largely inefficient due to the probabilistic population of different types of excitonic complexes, overall degrading the properties of quantum light emitters. Recent advancements in quantum optics theory and experimental settings marked a paradigm change towards resonant excitation methods to maximize QD benefits, especially valuable if combined with a QD site-control capability.

By exploiting resonant methods, population inversion of exciton and biexciton states can be achieved via Rabi oscillations[4,5], adiabatic rapid passage[6,7] and, later proposed and demonstrated, via robust schemes utilising phonon dressed states[8–10]. Irrespective of being a coherent or an incoherent phonon-assisted process, two-photon excitation is becoming a standard initialization process for the biexciton state in experiments and applications requiring a high quality pair of polarization-entangled photons. Near unity initialization fidelity[11], spectral purity[12], near perfect single photon emission[13] and others, sometimes in combination with recent integrated photonics advancements, are all critical outcomes of these approaches.

Experimentally the two-photon excitation scheme can be realized because of the non-degeneracy of the exciton and biexciton states arising from the interplay of Coulomb interactions and zero-dimensional geometrical constrictions[14]. The energy difference $\Delta E_B = E_X - E_{XX}$ between the exciton $E_X$ and biexciton $E_{XX}$ is referred to as the binding energy of the biexciton complex. In III-V epitaxial QDs, the $\Delta E_B$ value is typically within the range of a few meV. While coherent two-photon excitation (TPE) fulfils the energy conservation requirement $2E_{laser} = E_{XX} + E_X$, the laser photon energy $E_{laser}$ will be detuned from the exciton by $-\Delta E_B/2$ (Fig. 1a). Thus picosecond pulses satisfying the TPE condition can relatively easily address the biexciton state without, for example, the need for cross-polarization filtering to suppress the laser light which would compromise the polarization state, source intensity, purity and deterministic initialization.

Noteworthy, the majority of recent reports found in the literature deal with the so called binding biexciton case ($\Delta E_B > 0$). In a basic picture, higher order excitonic complexes in bulk, two- and one-dimensional structures are always binding, as the binding energy is the result of Coulomb interactions, and the stable complexes can appear only at lower energy[15]. On the other hand, in zero-dimensional structures, geometrical constraints limit the space available for the charge carriers to redistribute due to Coulomb interaction[14,16]. In specific cases, this can give rise to higher order complexes, so called antibinding biexcitons, with $\Delta E_B < 0$. Such states have been observed in a number of cases[15,17–21], however, no experimental study concentrating on the specifics of the antibinding state resonant initialization has been reported so far.

To address an antibinding biexciton by a resonant two-photon excitation, the laser energy has to be detuned positively from an exciton by $\Delta E_B/2$ (Fig. 1b). As we will show later, this can lead to a very different QD population dynamics which, unfortunately, might not necessary produce a favourable outcome for QDs with an antibinding biexciton in the contest of quantum information processing. Schematically this population and recombination complexity is outlined in Fig. 1b – in parallel to the expected two-photon excitation

a)Electronic mail: gediminas.juska@tyndall.ie

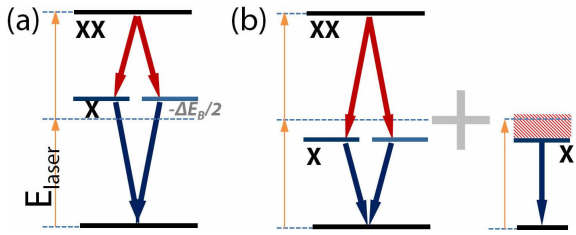

FIG. 1. Two-photon excitation schema in QDs with different types of biexcitons (XX). (a) A binding biexciton two-photon excitation (TPE). (b) An antibinding biexciton TPE competing with phonon-assisted exciton (X) population.

of a biexciton, favoured conditions to excite an exciton state through the interaction with acoustic phonons can occur, i.e. a single photon can give some of its energy to acoustic phonons and utilise the rest to excite only the exciton state.

In our manuscript, we are reporting systematic experimental evidence of this complex dynamics. As a case study, we have selected an $In_{0.25}Ga_{0.75}As/GaAs$ site-controlled Pyramidal QD system. All *single* QDs from this QD family exceptionally possess antibinding biexcitons[22] making them an interesting playground for fundamental research of resonant excitation phenomena. Noteworthy, when two or more identical single $In_{0.25}Ga_{0.75}As$ QDs are stacked with a separator of ⩽2 nm, they form interacting or coupled-QD complexes with very different energetic structure[23]. Some of them are single QD-like, with a binding or even degenerate biexciton[24], however, due to small binding energy (0.53±0.47 meV) and higher fine structure splitting values, these QDs are less attractive candidates for resonant TPE experiments.

The QDs have been prepared by metalorganic vapour phase epitaxy (MOVPE) on (111)B oriented GaAs substrates pre-patterned with 7.5 $\mu m$ tetrahedrons ensuring site-control with the precision of a few nanometers – arguably an important advantage over self-assembled QD systems. $In_xGa_{1-x}As$ QDs with a nominal Indium content of 0.25 were confined by GaAs (see Supplementary Material for more details). The most notable and attractive feature of these QDs is their high symmetry. An intrinsically highly symmetric ($C_{3V}$ in group theory terms) carrier confinement potential, theoretically foreseen for all (111) direction grown QDs[25,26] ensures a record density of QDs emitting polarization-entangled photon pairs under non-resonant optical[27] and electrical[28] excitation.

QDs have been characterised in a micro-photoluminescence set-up at 8 K in a closed-cycle helium cryocooler. Fourier transform limited 10 ps pulses have been pulse-shaped in a 4f (f=0.5 m) configuration with a portable tuneable-width slit. The base 80MHz 120 fs pulse-width emission was from a tuneable Ti:Sapphire laser. The laser emission was filtered with a series of 3 volume Bragg gratings utilised as narrow-band (0.3 nm) notch filters. Spectra have been taken by a CCD paired with a 1 m length monochromator with a $950 mm^{-1}$ groove density ruled diffraction grating. The second-order correlation measurements[18] were taken by filtering exciton and biexciton transitions with two identical monochromators and utilising single-photon avalanche photo detectors with dark counts of around 40 per second.

Figure 2a shows a QD spectrum ($\Delta E_B = -2.5\ meV$) obtained under two-photon excitation tuned resonantly to the biexciton state (several other representative QD spectra are shown in Supplementary Material). The observation of the biexciton peak confirmed the possibility to excite QDs with an antibinding state resonantly. Autocorrelation of both, an exciton (X) and biexciton (XX), transitions showed clear antibunching in the second-order (intensity) correlation curves confirming high quality single-photon emission from both lines. The measured values of $g^{(2)}(0)$ in the representative case (Fig. 2a insets) were rather low, i.e. 0.03±0.01 (X) and 0.015±0.015 (XX).

Interestingly, we systematically observed clear asymmetries in the intensity of X and XX transitions indicating substantially different population probabilities of the two states – a rather unexpected result for a state recombining in a cascade event. To understand the origins, we took the excitation power-dependent measurements presented in Fig. 2b. Characteristic Rabi oscillations of the excited level (XX) population were observed proving the coherent nature of the process. We attribute the damping to typical dephasing mechanisms, such as dephasing due to phonons at high excitation power[8]. As damping processes typically stabilise the population around 0.5, we estimate that at a $\pi$ area pulse the maximum achieved population probability of the biexciton state is 0.65, where the remaining 0.35 population is because of the phonon-assisted exciton initialization discussed below. Small laser energy detuning of +0.175 meV and -0.185 meV were sufficient to eliminate the coherent population and shift it towards a phonon-assisted excitation of the biexciton state process (see Supplementary Material, Fig. 3S).

The exciton, on the other hand, showed a different dependence. While a small oscillation correlated with the biexciton one was observed, an overall emission intensity growth was obtained in the whole power range. This can be predicted in view of the complex mechanism expected for populating an antibinding biexciton case. Specifically, two phenomena are contributing to the intensity of the exciton – 1) photons coming from the biexciton recombination cascade, and 2) photons coming from the exciton state which is excited non-resonantly and incoherently[9] through the acoustic phonon bath. The measured intensity is the integrated signal of both. Clear antibunching in the second-order correlation curve suggests that both are taking place during different excitation cycles. To discriminate the contribution of each, we assume for simplicity that the exciton population from the cascade is equal to the population of the biexciton. The subtracted result would represent the phonon-assisted X population showing monotonic growth. This is in a good agreement with theoretically expected and experimentally observed results for an exciton-only excitation[8,9]. To achieve high population values, this incoherent population process through a phonon-dressed state requires achieving a trade-off between the laser electric field strength and the state preparation time defined by the laser pulse length[29]. This scheme can provide a very robust way of state preparation once realised. The general principle applies for both, the exciton and biexciton, states (in Supplementary

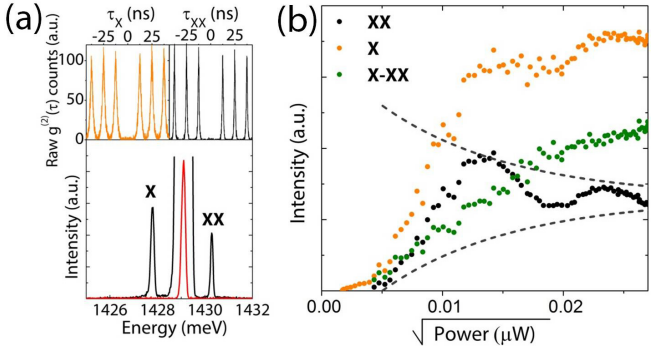

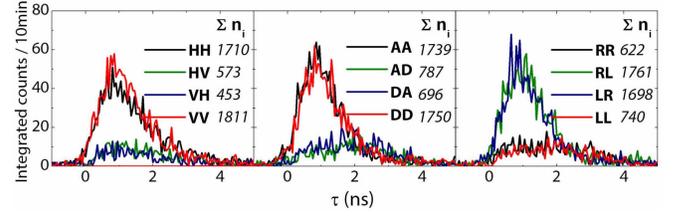

FIG. 2. (a) The spectrum under TPE excitation with 10 ps pulse (the red curve). The insets show the second order auto-correlation curves of the exciton (X) and biexciton (XX) transitions. (b) Excitation power dependence. The dashed lines are the envelopes of the damped Rabi oscillations. X-XX curve is the difference between the exciton and biexciton intensity at the same excitation power. These are the photoluminescence events occurring due to the phonon-assisted exciton population and which do not originate in the biexciton-exciton recombination cascade.

Material, Fig. 4S is showing phonon-assisted population as a function of laser detuning).

The second-order auto-correlation measurements evidently reveal that the two QD excitation scenarios are competing and do not occur within the same excitation cycle ($g^{(2)}_X(0) = 0.03\pm0.01$). It is an expected result as, with a probability dependent on the excitation conditions, a QD is excited either to the exciton or biexciton state – the population of one state blocks the population of the other. Provided that both processes can occur within the timescale of the laser pulse length in general, reexcitation of a QD becomes unlikely and results in low $g^{(2)}(0)$ values of the exciton state. While such two-channel excitation-recombination mechanism creates deterministic state preparation issues, it does not prevent the detection of photons originated solely during the biexciton recombination cascade. These photons are of primary interest due to the expected polarization entanglement.

To measure polarization entanglement of these pairs of photons, the biexciton photon was used to herald the exciton photon originated in the recombination cascade. By correlating them, second-order cross-correlation curves were obtained. Twelve polarization-resolved measurements have been made to obtain full sets of two-photon polarization states in linear, diagonal and circular bases (Fig. 3). Every individual curve has been integrated for the same duration, i.e. 10 minutes, to ensure a valid data processing procedure as the raw detection events triggered only by the same excitation pulse have been considered ($\pm6$ ns range). Raw detection counts (given as insets in Fig. 3) were used as the two-photon polarization state intensity (see Supplementary Material for the full procedure description).

The biexciton-exciton photon pair entanglement has been estimated by calculating the fidelity to the expected, maximally entangled Bell's state $\frac{1}{\sqrt{2}}(|HH\rangle+|VV\rangle)$:

FIG. 3. Polarization entanglement between the photons emitted during the biexciton-exciton recombination cascade. Full measurements in linear, diagonal and circular polarization bases. The fine-structure splitting of the representative QD is $0.4\pm0.4$ $\mu eV$.

$$f = \frac{1}{4}(1 + S_{33} + S_{11} - S_{22} + S_{30} + S_{03}), \quad (1)$$

where $S_{ii}$ are the two-photon Stokes parameters[30]. If the QD source is ideal, non-polarized, $S_{30} = S_{03} = 0$, and $S_{33}$, $S_{11}$, $S_{22}$ conventionally represent degrees of correlation in linear, diagonal and circular bases, respectively[31]. In our representative case, we obtained $S_{33} = 0.549$, $S_{11} = 0.403$, $S_{22} = -0.435$, $S_{30} = 0.004$ and $S_{03} = -0.049$, yielding the fidelity value $f = 0.586\pm0.004$. The value exceeding 0.5 is a clear indicator of entanglement ($f\in[0,1]$), even if the absolute fidelity value is far from recently achieved records[32]. The main entanglement degradation reason can be attributed to the usual cross-dephasing events which destroy the recombination cascade coherence. Indeed, a QD with a very small fine-structure splitting ($0.4\pm0.4$ $\mu eV$) and pure single photon emission, ideally, should emit pairs with a higher degree of entanglement. However, we systematically observed that these QDs under resonant excitation, in contrast to other type of QDs, maintain a relatively slow (0.78 ns in this specific case) exciton lifetime (see Supplementary Material part on the lifetime extraction), which is sufficient to increase a cross-dephasing probability substantially[33]. One of the most prominent decoherence mechanisms can be attributed to processing-induced defect states within the bandgap, which can be exited even with the laser tuned to the TPE resonance. The presence of such charge trapping states and destructive effect on optical properties of QDs from the same sample have been demonstrated previously[22]. In agreement to this, a significant linewidth broadening – by spectral wandering – of the exciton (160 $\mu eV$) and biexciton (117 $\mu eV$) in the here discussed representative case even under TPE proves the existence of a dynamically charged QD environment. Moreover, and in general, under non-resonant excitation conditions, these trapping states are also the main reason of poor single photon emission, linewidth broadening and QD charging. Thus results obtained under non-resonant excitation appeared of little interest in the context of this work, and they were not taken for comparison.

Systematic polarization resolved measurements revealed that the polarization states of photons emitted due to direct phonon-assisted excitation are strongly dependent on the polarization state of the laser. While being the matter of a broader dedicated study in the future, our preliminary observations suggest that the laser polarization state can be mapped on the polarization state of the exciton photon depending on the QD's energetic structure. A demonstration of the phenomenon using the same representative QD as above is shown in Fig. 4. TPE is typically performed using linearly polarized laser to overcome Pauli exclusion principle[5] (in Supplementary Material, we show that the biexciton excitation channel can be effectively suppressed with circularly copolarized photons). Such excitation can efficiently excite the conventional biexciton state, which then recombines in a cascade. The XX curve in Fig. 4 is representing these events. The distribution of linear polarization components here has been analyzed by rotating a half-waveplate placed in front of a fixed polarizer at the entrance of the monochromator. Small modulation in the XX intensity is characteristic to this QD and is in good agreement with the two-photon Stokes parameters' measurements showing that this QD source is slightly polarized. We tentatively attribute this polarization dependence to a weak mixing of heavy and light holes[34] which can vary in different QDs. The intensity of X events exceeding XX can be assigned to the phonon-assisted initialization of the exciton state (which will refer to as X-XX, i.e. X minus XX). In the case depicted in Fig. 4, these events are fully linearly polarized. The analysed laser intensity is also plotted to show that the polarization state of X-XX photons is nearly identical to the laser's one. The second-order auto-correlation measurement of the X transition (the antibunching $g^{(2)}(\tau)$ curve shown in the inset of Fig. 4) clearly confirmed that the source is fully described by a sub-Poissonian statistics, and it is not laser-related. By rotating the laser polarization by 45 deg (not shown), the X-XX maximum also shifted, with a just slightly different phase offset than the one visible in Fig. 4, confirming the existence of a correlation function between the polarization states of the laser and X photons created by phonon-assisted excitation.

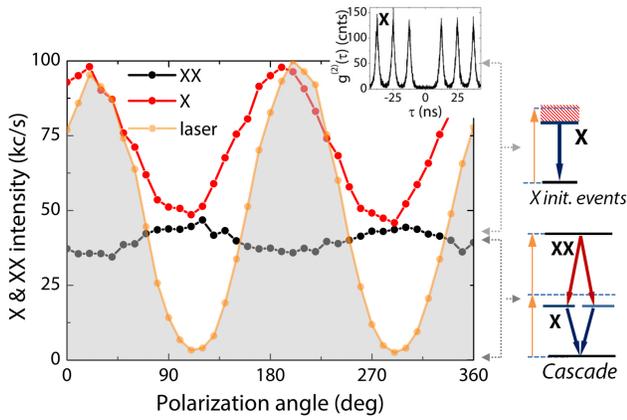

FIG. 4. Linear polarization analysis of the detected light originating from the laser, exciton and biexciton. An inset shows an antibunching curve collected from the exciton transition.

We speculate that the existence of the polarization phase offset between the maximum of the laser and X-XX photons is related to the energetic fine-structure of a QD. A possible scenario suggests that during the excitation moment the laser polarization state is mapped on to the Bloch sphere representing the exciton spin state[35–37]. In an ideally symmetric QD without the fine-structure splitting, the laser state should be mapped to the polarization state of the emitted photon. However, in the presence of a fine-structure splitting, the exciton spin state, depending on the initial state, can start to precess. The precession mapped on the exciton photon polarization state can be in principle resolved by fast single-photon detectors, which unfortunately we do not have access to. In our experiment we have been integrating emission events without time resolution. A slow precession of the exciton state could be used to explain the observed offsets between the maximum of the laser and X-XX photon events. This scenario should be tested by tuning the fine-structure splitting of the same QD (for example by strain[38]) and performing the same static and time-resolved polarization sensitive measurements.

While it might seem that an antibinding biexciton is a fundamental limitation in the resonant TPE approach, as true deterministic generation of non-classical light becomes compromised, it is worth pointing to a single reported case in the literature[39] which suggests that under certain conditions the TPE process actually can be the only excitation phenomenon (no phonon-assisted excitons) despite an antibinding biexciton. It is possible indeed that the TPE process is strongly affected by QD coupling to a cavity or it is highly dependent on the excitation conditions, as in ref.[39] the authors do not discuss nor observe any issue with the excitonic configuration. A known phenomenon of a strongly reduced excitation power needed to populate cavity-coupled QDs has been observed in the same type of devices (200 times reduction in comparison to QDs in a slab)[40]. It is possible that a weak laser field strength, which, in general, is driving an exciton coupling to the acoustic phonon bath, is not sufficient to populate the exciton state efficiently during the few picoseconds of the excitation. To achieve significant phonon-assisted exciton state population with a weak laser field, long preparation times (pulse duration), far exceeding the demonstrated very short lifetime (66 ps) of the biexciton, would be required[29,41]. That kind of microcavity-based devices driven by short (a few ps) pulses, in principle, are expected to be robust against phonon-assisted excitation under TPE conditions. Experimental evidence, even though with a binding biexciton, show that a reduced laser pulse length (from 13 ps to 7 ps) does not affect state population by resonant TPE, however, short pulses are less efficient in phonon-induced biexciton state population[42]. We, unfortunately, could not reliably verify this in our experiment due to a rather small binding energy ( -2 meV) and the high excitation powers needed, which were causing problems due to stray laser light.

In summary, we have demonstrated the population of site-controlled Pyramidal QD biexciton states by two-photon excitation. High quality single photon emission and good polarization entanglement between the photons emitted through the biexciton recombination cascade were measured. In all here presented cases we have been dealing with antibinding biexcitons. We showed that this specific fine-structure QD configuration leads to a complex excitation-recombination scenario. It was found that the laser tuned close to a two-photon resonance could populate the exciton state through the acoustic phonon side band. The latter excitation is less significant at low excitation powers, specifically at the exact resonance condition with a $\pi$ area pulse. However, as the two phenomena are competing, the true deterministic photon emission is compromised, at least in our $In_{0.25}Ga_{0.75}As$ QD system. On the other hand, there is no obvious reason why this phenomenon should not be observed in the other types of epitaxial QDs under similar, and thus at the moment typical, excitation conditions. The community members aiming to utilise QDs with antibinding biexcitons should be taking into account the findings presented in this work. To exploit the full potential of our highly symmetric site-controlled QDs, a different resonant excitation approach or conditions might be needed. Among possible candidates are phonon-assisted two-photon excitation energetically tuned above the biexciton transition[29] (the feasibility of this experiment is shown in Supplementary Material, Fig. 4S), and a variation of dichromatic wavelength excitation[43,44]. Also, a comprehensive study of different excitation conditions, such as pulse length, pulse area, and the effect of a microcavity could provide the solution to the compromised deterministic biexciton initialization. A here highlighted case presented by Wang et al[39] is in an agreement with the theory of phonon-assisted state preparation[29,41], both suggesting that phonon-induced population can be suppressed or minimised by designing devices that can be driven by low power and short (a few picoseconds) pulses. Finally, we observed strong correlations between polarization states of the laser and the exciton populated directly with phonon assistance. Our preliminary findings show that the laser state can be mapped on the exciton photon, and will be the subject of further studies.

## SUPPLEMENTARY MATERIAL

See supplementary material for sample fabrication and other resonant optical measurements details.

## ACKNOWLEDGMENTS

We thank Rinaldo Trotta for invaluable discussions on the experimental aspects. This research was supported by Science Foundation Ireland under Grant Nos. 15/IA/2864, 12/RC/2276-P2 and SFI-18/SIRG/5526.

## AIP PUBLISHING DATA SHARING POLICY

The data that support the findings of this study are available from the corresponding author upon reasonable request.

# Supplementary Material

## Biexciton Initialization by Two-Photon Excitation in Site-Controlled Quantum Dots: the Complexity of the Antibinding State Case

Gediminas Juska, Iman Ranjbar Jahromi, Francesco Mattana, Simone Varo, Valeria Dimastrodonato, Emanuele Pelucchi

*Tyndall National Institute, University College Cork, Lee Maltings, Dyke Parade, Cork, T12R5CP, Ireland*

Corresponding author: gediminas.juska@tyndall.ie

## Sample structure

The quantum dots (QDs) are grown by metalorganic vapour phase epitaxy (MOVPE) on (111)B oriented GaAs substrates pre-patterned with 7.5 µm pitch tetrahedral recesses (Fig. 1S). The exposed walls of the recesses are (111)A. Several epitaxial layers of $Al_xGa_{1-x}As$, GaAs and $In_xGa_{1-x}As(N)$ performing different function are grown at 730°C (thermocouple reading). For example, $Al_{0.8}Ga_{0.2}As$ acts as a wet-etching stop layer which enables selective substrate removal for the measurements of the pyramids in an apex-up geometry – a configuration which enhances light extraction by 2 or 3 orders of magnitude, and which was used in this work. The QD layer is confined by outer $Al_{0.45}Ga_{0.55}As$ and inner GaAs confinement barriers. The QD layer is $In_{0.25}Ga_{0.75}As(N)$ – during the growth the layer is exposed unsymmetrical dimethylhydrazine, which in this case appears to act as a surfactant improving QD symmetry. All the given concentration and thickness values are nominal, as the growth dynamics on non-planar surfaces is very complex[1]. During the complex growth process, a single QD is formed at the centre of every pyramid. The active layer material also forms other nanostructures (3 lateral quantum wires and 3 lateral quantum wells) interconnected with a QD.

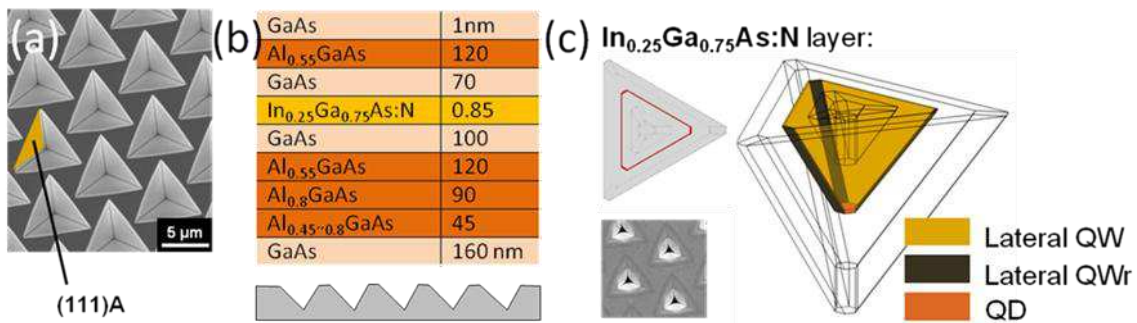

**Figure 1S.** (a) Patterned (111)B oriented GaAs substrate. (b) A stack of epitaxial layers grown on the non-planar patterned substrate. (c) An ensemble of nanostructures which forms inside the pyramidal recess.

## Other representative QD cases of TPE

Fig. 2S(a) shows three other representative QDs excited by two-photon excitation (TPE). In all measured QDs we have observed the same phenomena taking place, i.e. phonon-assisted excitation competing with the biexciton-exciton recombination cascade, as can be seen from the always brighter exciton (X) transition. Fig. 2S(b) presents linear polarization analysis of the exciton and biexciton (XX) emission. In a very good agreement with our findings presented in the main text, we see fully linearly polarized emission of the exciton part originating from the phonon-assisted excitation. Noteworthy, the biexciton intensity, in comparison to the slightly polarized case presented in the manuscript, is fully isotropic suggesting that no intrinsic phenomena, such as heavy-light hole mixing, are taking place[2,3].

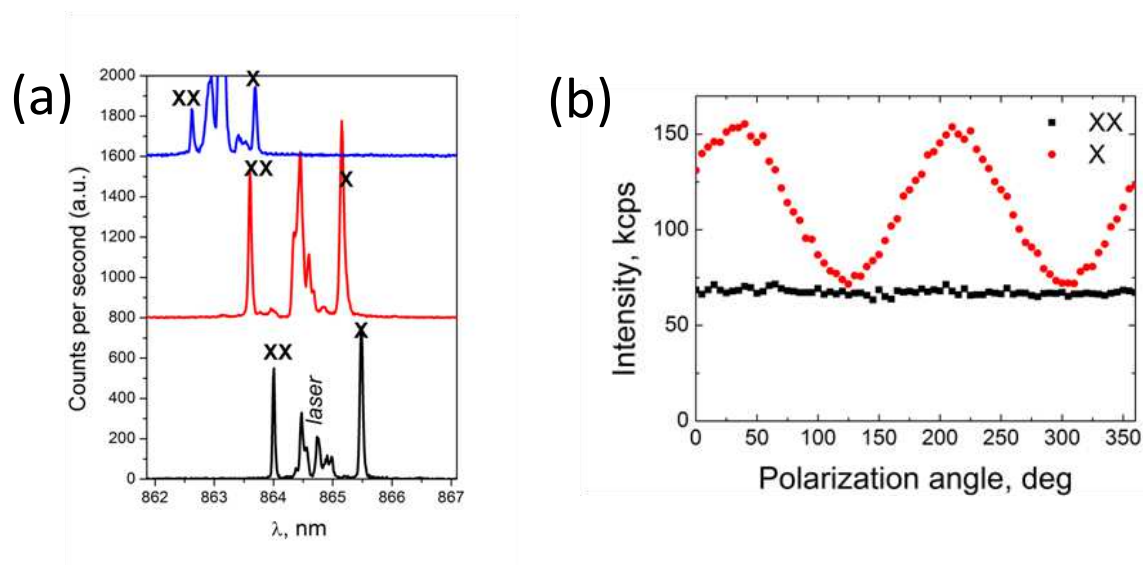

**Figure 2S.** (a) Spectra of 3 different QDs under TPE excitation. (b) Linear polarization analysis of X and XX transitions emitted from a different QD showing well reproducible behaviour presented in the main text.

## Off-resonance excitation

Fig. 3S shows power dependence of the biexciton (XX) and phonon-assisted (X-XX) exciton when the laser is tuned precisely to the two-photon resonance and off-resonance, i.e. positively and negatively detuned by +0.175 meV and -0.185 meV, respectively. As can be seen, the coherent nature of the state initialization characterized by Rabi oscillations is compromised, and the biexciton initialization, in analogy to the exciton case, is expected to be induced by phonons[4,5]. X-XX intensity changes can be attributed to different phonon density distribution. Fig 3S(b) shows X-XX under the resonant TPE condition plotted as a function of power (the logarithmic scale helps to visualize the initial part of the power dependence).

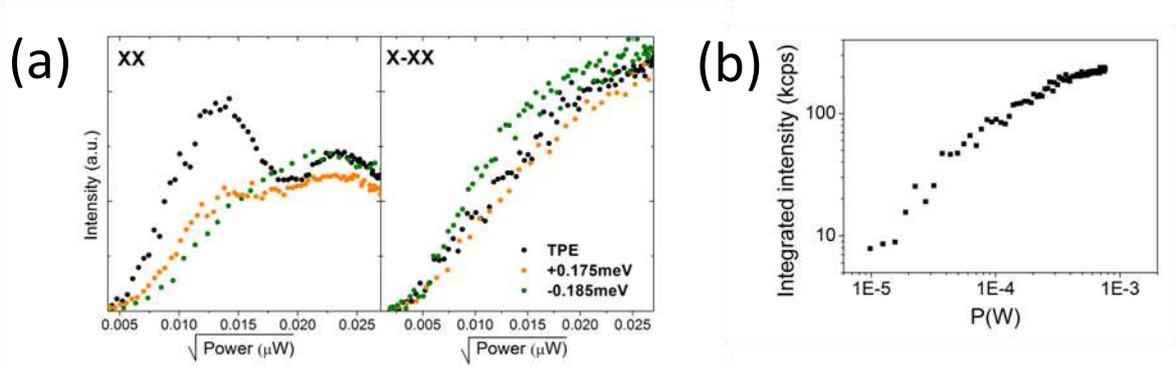

**Figure 3S.** (a) Power dependence of the biexciton and phonon-assisted exciton emission under different laser energy tuning conditions. (b) Phonon-assisted exciton under resonant TPE as a function of power.

Fig. 4S shows the exciton intensity dependence on the laser energy detuning. We stress that the exciton here fully originates from the biexciton recombination cascade. This was possible because, as we show in Fig. 2S(b), the phonon-assisted exciton is fully linearly polarized, and thus can be fully blocked with a cross-polarized polarizer. As the biexciton (and the subsequently recombining exciton) intensity is isotropic, such configuration will fully block phonon-induced excitons, however will transmit half of the light originating from the cascade. This study can give a good insight on the phonon-assisted TPE process.

The motivation for this study is the following. The competition of the two QD population phenomena is not favourable for practical applications as true deterministic generation of non-classical light becomes compromised. A possible solution to this problem has been proposed by Barth *et al*. (ref. 5). Large area pulses (>20π) positively detuned to the respect of both, the exciton and biexciton, are expected to excite the biexciton state with a near unity probability. Our experiment with ~π pulses detuned positively from the biexciton (Fig. 5S) shows that we can excite the exciton (and biexciton) state even with big (~2.5 meV) detuning values which in principle would allow to test the mentioned idea above. However, due to the experimental limitations (specifically, the power loss in the 4f pulse shaping set-up and laser light filtering issues) we could not achieve the required power levels. The results serve as a good indicator that such experiment is feasible and can be implemented with different experimental approaches.

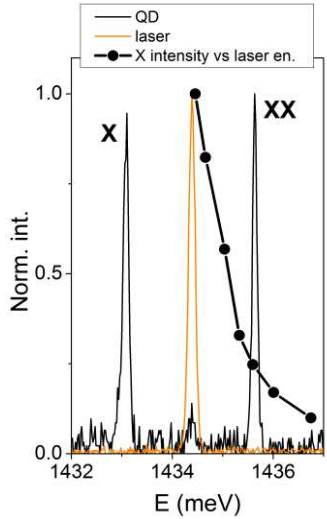

**Figure 4S.** The exciton intensity dependence as a function of a laser pulse energy detuning. The spectrum is linearly polarized in a way to transmit photons originating from the biexciton recombination cascade only.

## Biexciton excitation channel suppression

We checked also another issue – the suppression of the biexciton recombination cascade which should be achieved utilising circularly polarized excitation light (Fig. 5S(a)). Because of the Pauli exclusion principle, TPE processes of the biexciton by circularly co-polarized photons are fundamentally forbidden. We indeed observe a strong suppression of the biexciton state. The experimentally observed weak signal from the biexciton in Fig. 5S(a) can be attributed to a non-perfect preparation of the laser polarization state. Otherwise, in an ideal case, the laser light couples only to the exciton state. As the electron-phonon interaction is spin independent, the polarization state of the laser is initially mapped on the Bloch sphere of the exciton, which, in the presence of a fine-structure splitting, starts to precess. The inset of Fig. 5S(a) shows linear polarization behaviour for the exciton photons. The non-uniform distribution of linear polarization components indicates that the polarization state of the ensemble of photons is not purely circularly polarized due to the FSS of a few µeV in that specific dot. Fig. 5S(b) shows non-polarized spectra obtained under circularly and linearly polarized excitation.

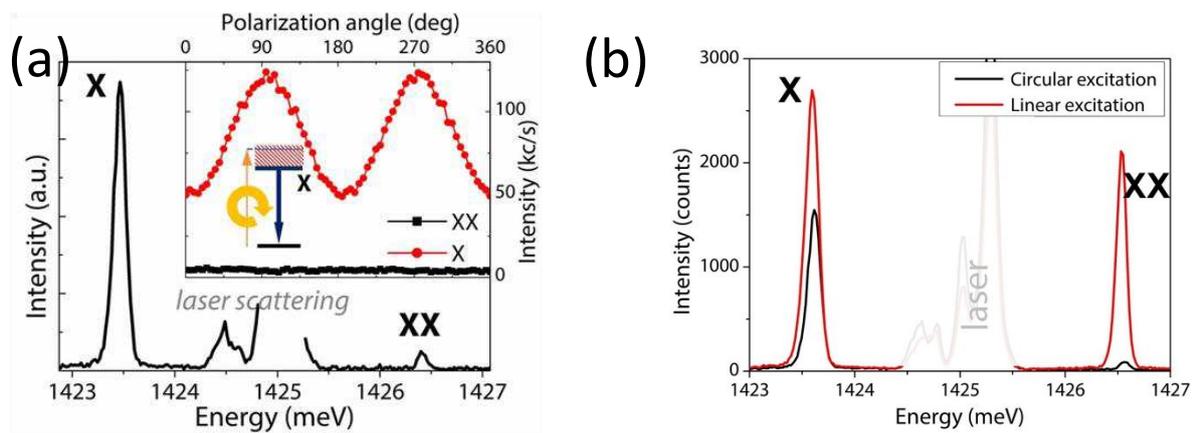

**Figure 5S.** (a) QD photoluminescence under circularly polarized TPE showing strong suppression of the biexciton recombination cascade. Here a different QD was used than the one discussed in the main text. (b) Comparison of non-polarized spectra obtained under circularly and linearly polarized laser light tuned to the TPE resonance (π pulse area for the biexciton transition).

# Processing of the second-order correlation curves:

1. **Polarization-resolved**

Fig. 6S(a) shows linear polarization analysis of the exciton and biexciton transition. As we discuss in the main manuscript, exciton photons created by phonon-assisted excitation are strongly polarized as the laser polarization state is directly mapped onto the spin state of the exciton. Being strongly polarized, phonon-assisted events invalidate the conventional normalization procedure to calculate the second-order correlation function value as an intensity of the projected two-photon polarization state. For example, Fig. 6S(b) shows a comparison of correlation curves between horizontally and vertically co-polarized biexciton and exciton photons. While the $g^{(2)}(0)$ peak is composed of events originating only from the biexciton and exciton recombination cascade, the side (uncorrelated) peaks arise due to contributions from all the events. As shown, horizontally polarized exciton photons originating from phonon-assisted excitation paths increase the side peak intensity nearly twice. Conventionally, when the source is not polarized, the two-photon polarization state intensity is obtained by normalizing the $g^{(2)}(0)$ peak using an average value of the side peak integrated intensity. However, this method is only valid when the source is not-polarized. To overcome this issue, every individual curve has been integrated for the same duration, i.e. 10 minutes, and raw detection counts of the $g^{(2)}(0)$ peak were used as the two-photon polarization state intensity. The comparison of the two-photon polarization state intensity obtained by the two methods is shown in the Table 1.

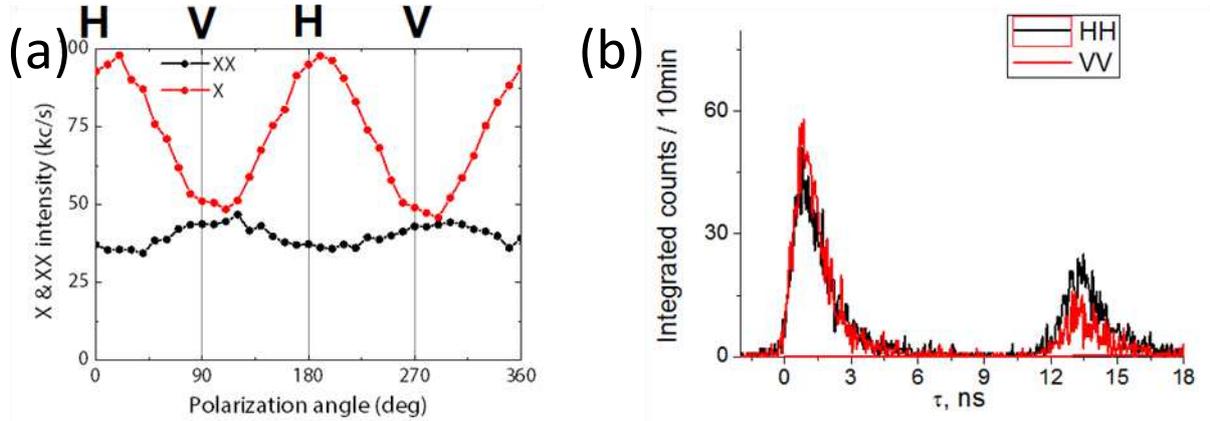

**Figure 6S.** Polarization-resolved $g^{(2)}(\tau)$ (a) Linear polarization analysis of exciton and biexciton transitions. (b) Comparison of correlation curves between horizontally and vertically co-polarized biexciton and exciton photons.

**Table 1.** Two-photon polarization state intensity: $g^{(2)}(0)$ vs. raw counts

|     | $g^{(2)}(0)$ | $\Sigma n_i$ |     | $g^{(2)}(0)$ | $\Sigma n_i$ |     | $g^{(2)}(0)$ | $\Sigma n_i$ |
| --- | --- | --- | --- | --- | --- | --- | --- | --- |
| HH  | 1.77 | 1710 | DD  | 2.47 | 1750 | RR  | 0.94 | 622  |
| HV  | 0.92 | 573  | DA  | 0.95 | 696  | RL  | 1.82 | 1761 |
| VV  | 3.65 | 1811 | AA  | 1.89 | 1739 | LL  | 0.81 | 740  |
| VH  | 0.53 | 453  | AD  | 1.04 | 787  | LR  | 2.22 | 1698 |
| Sum: | 6.87 | 4547 |     | 6.35 | 4972 |     | 5.79 | 4821 |

## 2. Lifetime extraction

The intrinsic lifetimes of the biexciton and exciton were extracted from the second order correlation measurements, by correlating two APD signals (the conventional Hanbury Brown-Twiss arrangement) and deconvoluting the corresponding state population function.

The extracted biexciton lifetime was found to be 0.44 ns in the representative case studied in the main manuscript. As the state is created coherently by the laser, we deconvoluted the measured instrument response function (IRF) – using the autocorrelation signal of the 10 ps pulse laser – from the correlation curve of the laser and the biexciton photons(Fig. 7S(a)).

The intrinsic exciton lifetime, on the other hand, cannot be extracted using similar correlations. While typically a lifetime trace is defined as a histogram of correlated events between a fast electric laser synchronisation signal (or an optical laser signal like in our case) and the transition of interest, such measurement not necessarily can directly provide the intrinsic lifetime of the state. In our case, the measured trace is the result of a convolution between the biexciton decay function and the exciton decay. After deconvoluting the biexciton state, we extracted an exciton lifetime of 0.77 ns (Fig. 7S(b)). However, in this particular case, extra care has to be taken because, due to the presence of phonon-assisted events, the exciton population function is either from the laser or from the biexciton decay.

To rule out the contribution of phonon-assisted events, the exciton lifetime can be reliably extracted from the $g^{(2)}(0)$ peak of the correlation curve obtained between the biexciton and exciton cascade. The $g^{(2)}(0)$ peak is composed of correlation events originating only from the biexciton-exciton recombination cascade events, and it is only convoluted with the IRF (Fig. 7S(c)). After the deconvolution procedure, we obtained the exciton lifetime of 0.78 ns, close enough to our previous result.

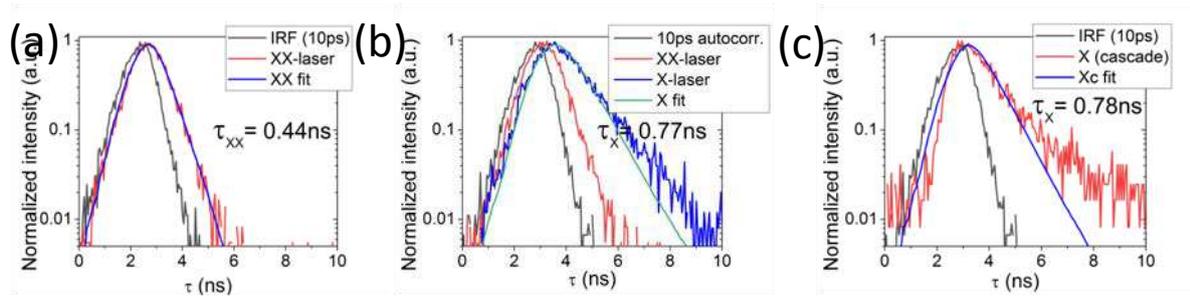

**Figure 7S.** Lifetime extraction. (a) The biexciton-laser correlation. (b) Correlations with the laser signal. (c) Exciton lifetime from the biexciton-exciton recombination cascade, i.e. the $g^{(2)}(0)$ peak.